\documentclass[aps,prb,preprint,superscriptaddress, showpacs]{revtex4}

\usepackage{graphicx}
\usepackage{bm}
\usepackage[dvips]{color}

\newcommand{\kspace}{\emph{k}-space}

\begin{document}

\preprint{}

\title{Plasmon scattering from single sub-wavelength holes}

\author{N. Rotenberg} \email{rotenberg@amolf.nl}
\author{M. Spasenovi\'{c}}
\author{T. L. Krijger}
\author{B. le Feber}
\affiliation{Center for Nanophotonics, FOM Institute AMOLF, Science Park 104, 1098 XG, Amsterdam, The Netherlands}
\author{F. J. Garc\'{\i}a de Abajo}
\affiliation{IQFR -- CSIC, Serrano 119, 28006 Madrid, Spain}
\author{L. Kuipers}
\affiliation{Center for Nanophotonics, FOM Institute AMOLF, Science Park 104, 1098 XG, Amsterdam, The Netherlands}

\date{\today}

\begin{abstract}
We map the complex electric fields associated with the scattering of surface plasmon polaritons by single sub-wavelength holes of different sizes in thick gold films.  We identify and quantify the different modes associated with this event, including a radial surface wave with an angularly isotropic amplitude. This wave is shown to arise from the out-of-plane electric dipole induced in the hole, and we quantify the corresponding polarizability, which is in excellent agreement with electromagnetic theory.  Time-resolved measurements reveal a time-delay of $38\pm18\,$fs between the surface plasmon polariton and the radial wave, which we attribute to the interaction with a broad hole resonance.
\end{abstract}


\maketitle


The interaction of light, and in particular of surface waves such as surface plasmon polaritons (SPPs), with sub-wavelength holes in metallic films leads to phenomena such as extraordinary optical transmission (EOT)~\cite{EOT_NatRev, EOT_Rev} and negative index metamaterials~\cite{Theory_NR_I, Exp_NR_I, Exp_NR_II, Exp_NR_III}.  In this context, Sub-wavelength holes are important in diverse fields of research, such as nanophotonics, enhanced nonlinear optics, and biosensing, and consequently, with a few notable exceptions~\cite{EOT_Exp_Lienau, EOT_Exp_Lalanne}, experiments focus on the macroscopic consequences of the light-hole interactions.  Conversely, much theoretical work has been devoted to their understanding from a microscopic viewpoint~\cite{Theory_Lalanne, Theory_GVidal, Theory_Liu, Tutorial_Abajo}.  That is, while the models intrinsically consider scattering of light and SSPs from sub-wavelength defects, often in terms of their polarizabilities, most experimental studies address collective response features such as EOT spectra and their dependence on changes in sample geometry, film thickness~\cite{Film_Thickness}, and hole shape~\cite{Hole_Shape}.  In essence, a detailed investigation of the dynamics of plasmonic scattering remains incomplete.

In this Letter, we present a systematic experimental study of the scattering of SPPs from individual circular sub-wavelength holes of different diameters in optically thick gold films.  We use phase-sensitive time-resolved near-field optical observations to measure the plasmonic scattering on a nanometer length- and femtosecond time-scale, determining the angularly resolved scattering amplitude for sub-wavelength holes of different sizes and quantifying the temporal dynamics of the scattering events.  We show that these results can be explained in terms of the polarizability of the hole, in excellent agreement with electromagnetic theory.

\begin{figure}[!h]
  \includegraphics[width=6cm]{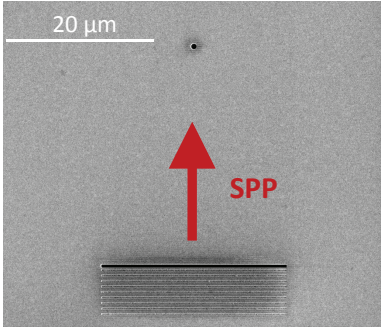}\\
  \caption{(Color online) Scanning electron micrograph of a SPP launcher and a hole.  The arrow shows the direction of propagation of the excited SPPs. Illumination is provided from underneath the film.}\label{fig:Sample}
\end{figure}

We investigate the scattering from individual holes, with diameters ranging from 50 to 980 nm, milled with a focused ion beam into a 200 nm thick gold film deposited on a glass substrate.  We launch SPPs on the air-gold interface by illuminating a slit in the film with either 150 femtosecond FWHM pulses or a continuous wave laser source centered at a wavelength of 1550\,nm.  At this wavelength, a waveguide mode is supported by holes with diameters above 760 nm. These SPPs are efficiently directed towards the hole by a Bragg grating that is milled into the film on the opposite side of the slit~\cite{Launcher}, as shown in Fig.~\ref{fig:Sample}.
We measure the complex electric field of this system, at a height of 20 nm, with near field microscopy~\cite{PHANTOM}.

\begin{figure*}[!htp]
  \includegraphics[width=16cm]{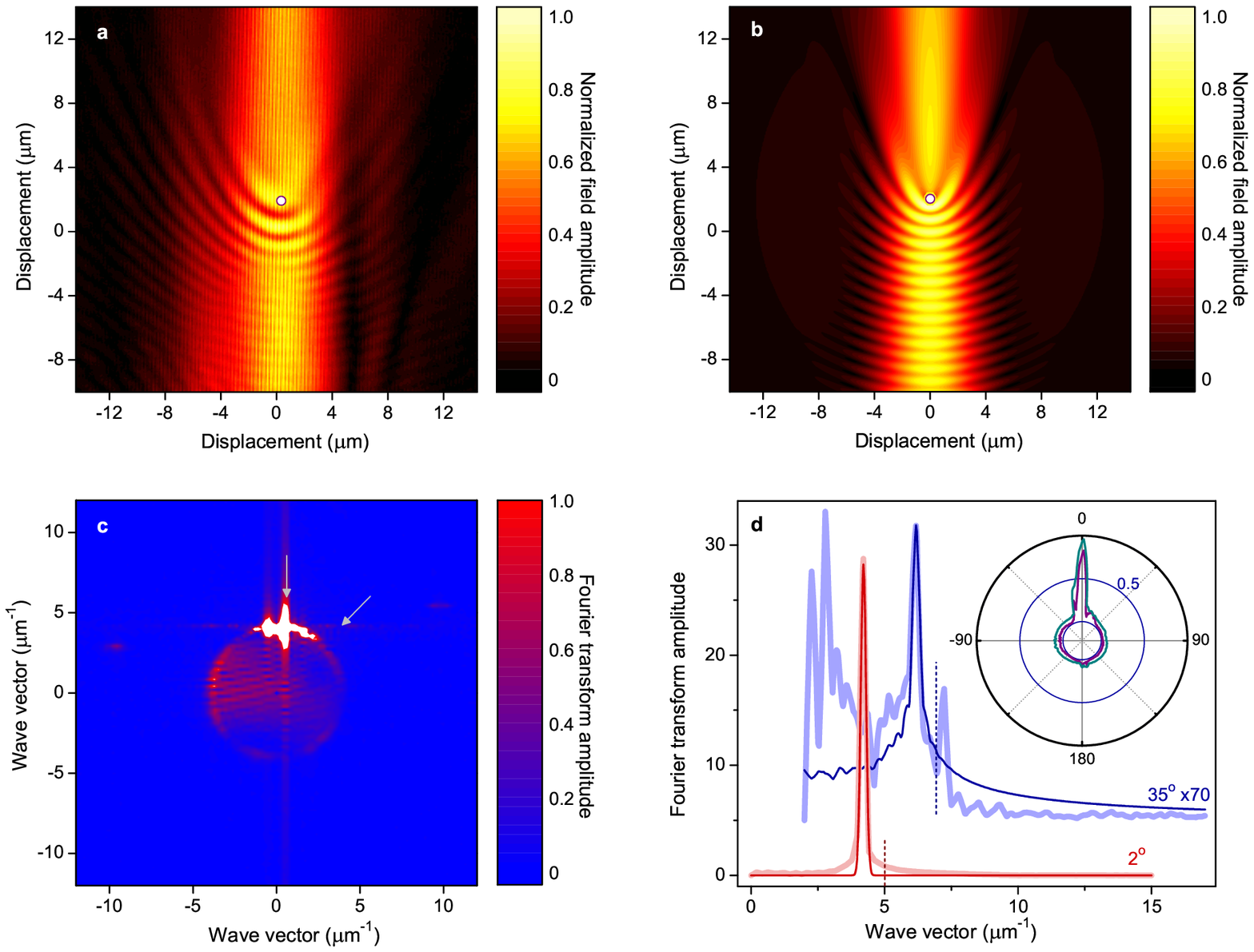}\\
  \caption{(Color online) Typical near-field measurements of plasmon scattering from a sub-wavelength hole.  \textbf{(a)} Real space image at the time when the peak of the SPP pulse is in the scan area.  The pulse has gaussian profile and is incident from the bottom of the image. The hole position is indicated by a solid circle. Fringes caused by interference between incident and scattered plasmons are clearly visible.  \textbf{(b)} Calculated electric field amplitude under the conditions of (a). \textbf{(c)} \kspace\ representation of the data shown in (a).  Incident and transmitted SPPs are associated to the vertical line, while scattered plasmons appear as a circle.  The image has been saturated to make the signature of scattered modes clearly visible.  \textbf{(d)} Cuts along $2^{\circ}$ and $35^{\circ}$ radial directions in \kspace\ [see arrows in (c)] showing signatures of forward and radial plasmon waves, respectively.  The $35^{\circ}$ line has been magnified by a factor of 70, its amplitude offset by 5, and position shifted by 2\,$\mu$m$^{-1}$ for clarity.  Thick and thin curves are the data and the corresponding fits, respectively. Inset: angle-dependent integrated field amplitude; the outer blue curve is integrated up to $k = 5$ $\mu$m$^{-1}$ (see dashed curves in main plot); the inner purple curve represents the combined amplitude of the two surface modes identified in (c).}\label{fig:Example}
\end{figure*}

In Fig.~\ref{fig:Example} we present typical experimental results for a hole with a diameter of 875 nm.
We observe fringes in the amplitude that originate from the interference of the forward propagating SPP with the scattered waves.  Here, the dark vertical stripes near 5 and 8 $\mu$m are attributed to geometrical imperfections in the area of the launching slit. The salient features of the data are reproduced by first-principle calculations [Fig.~\ref{fig:Example}(b)] performed using a modified version of a model presented elsewhere~\cite{Tutorial_Abajo}.  This method, which involves a rigorous expansion of Maxwell's equations, both in terms of plane waves on either side of the film and in cylindrical waves inside the hole, is used to calculate the electric and magnetic polarizabilities of the hole, which are in turn used to obtain the scattered fields~\cite{SI}.

The different waves present in the scattering event are readily identifiable in \kspace~\cite{kSpace} after a Fourier transform [Fig.~\ref{fig:Example}(c)].  In \kspace , signatures of the forward propagating SPP, evidenced by the peak near $2^{\circ}$, and the scattered wave, which emerges as a circular pattern, are readily separable.  These signatures are shown in Fig.~\ref{fig:Example}(d), in which both the data (thick shaded curves) and the fits for the different surface waves (thin dark curves) are presented.  The distributions of both waves peak around a wave vector $k = 4.2$ $\mu$m$^{-1}$, which is in good agreement with the theoretical value of $k_{sp} = k_0 \left(\epsilon / \left(\epsilon + 1\right) \right)^{0.5} = 4.1\,\mu$m$^{-1}$, obtained from the dielectric constant of gold at 1550 nm, $\epsilon = -115 + 11i$~\cite{AuNK}.

In the inset of Fig.~\ref{fig:Example}(d), we show the angular dependence of the in-plane electric field distribution.  Here, the outer curve represents the total integrated field amplitude while the inner curve represents the integrated amplitude that we can attribute to the surface waves. We find these curves by integrating up to $k = 5.0$ $\mu$m$^{-1}$ under the data and fits, respectively, in each direction (main part of the figure).

\begin{figure}
  \includegraphics[width=8.3cm]{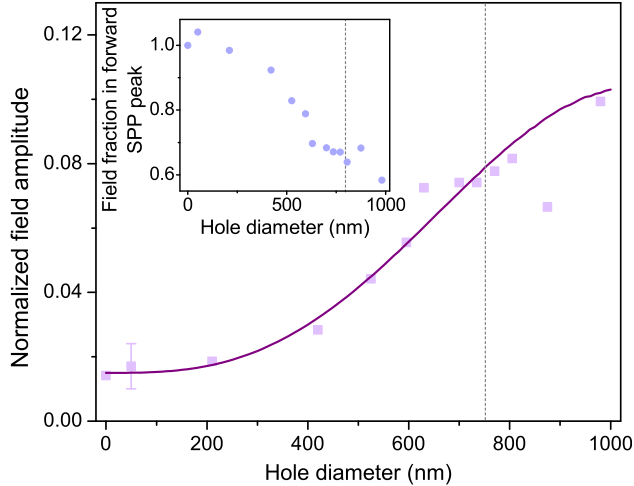}\\
  \caption{(Color online) Hole-size dependence of the plasmonic scattering. We plot the radially scattered field amplitude normalized to the peak of the forward propagating SPP as a function of hole size.  The solid curve shows the calculated hole-size dependence of $p_z$ (and hence the calculated scattered amplitude).  Inset: fraction of field contained in the forward propagating SPP peak, normalized to unity for a featureless film, as a function of hole size. The vertical dashed lines show the cutoff of the lowest-order guided mode in a deep hole for 1550 nm light.}\label{fig:Size}
\end{figure}

First, we observe no angular dependence of the radial mode and attribute the slight left-right asymmetry found in our measurements to either a slightly asymmetric hole or near-field tip. Second, for this relatively large hole the amplitude content of the radial wave is about 7$\%$ of the amplitude of the forward propagating SPP.  As we show below (Fig.~\ref{fig:Size}), for an 875 nm hole $\sim 0.7$ of the forward propagating SPP amplitude is contained in the incident beam, and hence the amplitude of the radial wave is $\sim 10\%$ that of the \emph{incident} SPP. Note that our incident Gaussian beam is $\sim 6 \, \mu$m width, which is larger than the hole (see~\cite{SI} for a discussion on the scattering cross section of the hole.  Finally, from this data we can also estimate that approximately half of the radially scattered field amplitude is in the surface modes.  The remaining half of the radial fields have in-plane wave vectors that are smaller than $k_0 = 2 \pi / \lambda = 4.05$ $\mu$m$^{-1}$, corresponding to radiative modes of light, which propagate upward, away from the metal surface.  Qualitatively, these are seen as the regions of non-zero amplitude within the surface wave circle in Fig.~\ref{fig:Example}(c), or in the difference between the inner and outer curves (other than near $0^{\circ}$) in Fig.~\ref{fig:Example}(d).

In order to determine the nature of the observed scattered field, we consider its asymptotic form~\cite{SI}
\begin{eqnarray}\label{eq:asymptE}
    \nonumber \lefteqn{\left(E_{x},E_{y},E_{z}\right)=2k_{0}^{2}\left(\frac{2\pi}{k_{sp}R}\right)^{\frac{1}{2}}\left(\frac{\epsilon}{1+\epsilon}\right)^{2}\left(\frac{1}{1-\epsilon}\right)} \\
    & & \, \times \, e^{-i\frac{\pi}{4}}e^{i\left(k_{sp}R+w_{sp}z\right)} A^{sc}\left(\varphi\right)\left(-w_{sp}\hat{R}+k_{sp}\hat{z}\right),
\end{eqnarray}
where the azimuthal scattering amplitude
\begin{equation}\label{eq:Asc}
    A^{sc}\left(\varphi\right)=-p_{x}\cos\varphi-\sqrt{1+\epsilon} \, m_{y}\cos\varphi+\sqrt{\epsilon} \, p_{z}
\end{equation}
is given for excitation by a SPP that travels along the $x$ direction.  Here, $R$ is the in-plane radial distance to the hole center, $z$ is the height above the film surface, $w_{sp} = -k_0 / \left(\epsilon+1\right)^{0.5}$ is the out-of-plane component of the SPP wave vector, $\hat{z}$ and $\hat{R}$ are out-of-plane and in-plane unit vectors, $p_x$ and $p_z$ are the in- and out-of plane electric dipoles, and $m_y$ is the in-plane magnetic dipole of the hole.  Consequently, the isotropic nature of the detected scattered field [inset to Fig.~\ref{fig:Example}(d)] suggests that we are mostly sensitive to the contribution from $p_z$.  This is not entirely unexpected since the contributions of other dipole components peak near $\varphi = 0^{\circ}$ and $180^{\circ}$ and are therefore hard to separate from the incident SPP field.  However, we note that to accurately reproduce the field in all directions [Fig.~\ref{fig:Example}(b)] the fields that arise due to $m_y$, as well as those due to $p_z$, are required~\cite{SI}.

\begin{figure}
  \includegraphics[width=8.3cm]{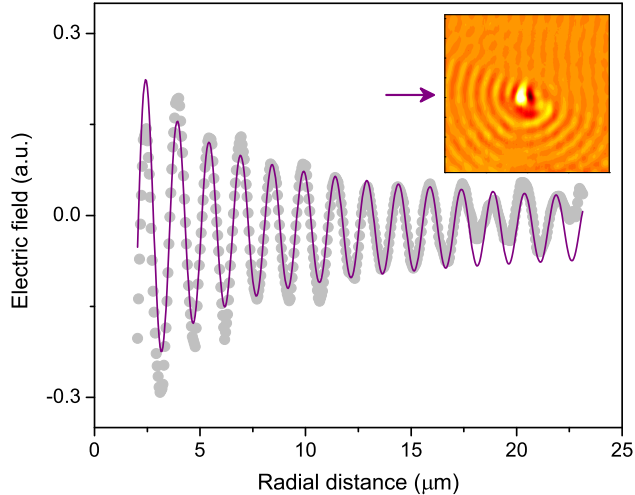}\\
  \caption{(Color online) Radial distribution of the field scattered by a 875 nm hole.  The measured real part of the electric field is shown by dots, while the curve is an $\exp{\left(i k R\right)}R^{-a}$ fit.  The inset shows the two-dimensional image of the real part of the electric field, and the arrow indicates the line along which the data shown in the main part of the figure is taken.}\label{fig:Radial}
\end{figure}

We image the $p_z$ dependent electric field by filtering out the forward propagating SPP mode in \kspace\ and then Fourier transforming back to real-space.  In Fig.~\ref{fig:Radial} we show this field, both near $\varphi = -90^{\circ}$ where only a contribution from $p_z$ is expected, and in a 20 $\mu$m$^2$ area around the hole (inset).
A fit of the form $\exp{\left(i k R\right)}R^{-a}$ reproduces the experimental data with $k = 4.20 \pm 0.02$ $\mu$m$^{-1}$ and $a = 0.57 \pm 0.10$. Hence, the measured amplitude damping is in excellent agreement with the $1/\sqrt{R}$ behavior predicted for circular waves by Eq.~(\ref{eq:asymptE}), as required for energy conservation for radial propagation on a plane. The discrepancy between the data and fit close to the hole may be indicative of the presence of creeping waves, which can have substantial amplitudes at short distances~\cite{Theory_Lalanne}. We can also use Eq.~(\ref{eq:asymptE}) to express the ratio of the scattered field to the incident field for $\varphi = 90^{\circ}$ in terms of the electric polarizability
\begin{equation}\label{eq:alphaE}
    \frac{A^{sc}}{A^{in}}
     \approx 2\sqrt{2\pi}\left(\frac{\epsilon}{1+\epsilon}\right)^{2}\left(\frac{\epsilon}{1-\epsilon}\right)k_{0}^{2}w_{sp}\alpha_{E}.
\end{equation}
Hence, we measure $\left|\alpha_E\right| = \left(4.1 \pm 0.1\right) \times 10^{-21}$ m$^3$, which is larger than, but comparable to, the value $\left|\alpha_E\right| = 3.3 \times 10^{-21}$ m$^3$ calculated for a hole in a perfectly conducting film~\cite{SI}.

\begin{figure}
  \includegraphics[width=8.3cm]{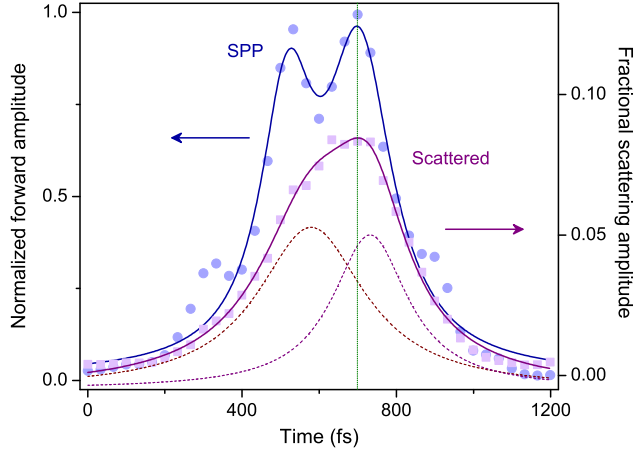}\\
  \caption{(Color online) Temporal dynamics of plasmon-hole scattering. The normalized forward field amplitude (circles, left axis) and the corresponding fractional scattered field (squares and diamonds, right axis) are shown as functions of the time delay.  The solid curves are double-peak fits which provide guides to the eye; the dashed curves show the individual peaks of the scattered amplitude fit; these lag behind the peaks of the forward amplitude (vertical dotted line at 700 fs, for the latter peak) by $\sim 40\,$fs.}\label{fig:Temporal}
\end{figure}

In Fig.~\ref{fig:Temporal} we present the temporal dynamics of plasmonic scattering, showing the amplitudes of both the forward and scattered radial modes as a function of time.
The forward amplitude is averaged over a range of $\pm 5^{\circ}$ while the scattered amplitude is averaged over a similar range in the transverse directions ($\theta = \pm 90^{\circ}$) to ensure no overlap between the modes. Both amplitudes are normalized to the peak of the forward propagating SPP, and the second peak of the double peak results from SPPs that are initially launched backward and subsequently reflect from the Bragg grating towards the hole~\cite{Launcher}. Most strikingly, we observe that the radial wave peaks one time-step \emph{after} the forward propagating SPP: there is a time-delay between the forward propagating SPP pulse and the radial pulse resulting from the plasmonic scattering.  By averaging over the time-delay that we measure between incident SPP and the scattered wave, in all available directions, we quantify this shift as $38 \pm 18 \,$fs. While the frequency dependence of the polarizability of the hole alone results in a shift only a few fs, it is able to explain the observed shift when combined with a modestly chirped incident pulse.  That is, since the different plane waves that form our pulse scatter from the hole with different amplitudes and phases, their reconstitution into a beam results in the observed time shift.  This shift, then, represents a first measurement of the complex spectral response of an individual hole.



We study the effect of hole size on the plasmonic scattering for hole sizes ranging from 50 to 980 nm. The results are summarized in Fig.~\ref{fig:Size}.
The inset depicts the fraction of the field amplitude that is contained in the forward propagating SPP mode, which we normalize to the field on films without a hole.  As expected, less energy is found in this plasmonic wave as the hole size increases and hence scatters more energy.

The main part of Fig.~\ref{fig:Size} depicts the dependence of  the scattered field amplitude of the radial wave normalized to that of the forward plasmon field (symbols) on the hole diameter, $d$.  This dependence is accurately reproduced by the calculated electric polarizability of the hole (curve)~\cite{SI}.  An offset of 0.017 is present in the data (and added in the calculations) due to measurement noise, and represents a signal-to-noise ratio of about 50 to 1, while the error bar is mainly indicative of the asymmetry of our measurements.  For the smaller holes, well below the cutoff, the field amplitude increases as $\sim d^3$, where the divergence from a perfect dipolar behavior is due to the finite size of the hole and the film thickness.  As the hole size increases, and in particular once the cutoff is reached and waveguide modes in the hole become accessible, the scattered amplitude begins to saturate. We attribute this behavior to an increase in energy flow through the hole~\cite{SI}.  Finally, we note that the scattered amplitude peaks at about 10$\%$ of the forward propagating SPP amplitude.


To summarize, we have captured the near-field interplay between the waves associated with plasmon scattering from sub-wavelength holes and mapped both their dependence on hole size and their temporal dynamics.  We show that both the hole size dependent amplitude and the time-delay of the scattered mode can be understood in terms of the calculated polarizability of the hole.  Our results provide a comprehensive understanding of hole-plasmon scattering and should, in the future, allow for the optimization of the interaction between holes. Consequently, the geometric properties of hole arrangements can be tuned to create electromagnetic hot spots. These can be used to, for example, enhance emission from dye molecules, address individual quantum emitters, or create near field distributions for sensing or imaging purposes.  Hence, these results have implications for a broad range of nanoplasmonic applications.

\section{Acknowledgment}
This work is part of the research program of the Stichting voor Fundamenteel Onderzoek der Materie (FOM), which is financially supported by the Nederlandse Organisatie voor Wetenschappelijk Onderzoek (NWO).  F.J.G. de A. acknowledges support from the Spanish MEC (MAT2010-14885 and Consolider NanoLight.es).

\end{document}